\documentclass[12pt]{article}
\usepackage{epsfig,graphicx}

\title{QED spectra in the path integral formalism}
\author{  Yu.A.Simonov,\\ Institute of Theoretical and Experimental
Physics\\ 117218, Moscow, B.Cheremushkinskaya 25, Russia}
\date{}

\newcommand{\be}{\begin{equation}}
\newcommand{\ee}{\end{equation}}

\def\fun#1#2{\lower3.6pt\vbox{\baselineskip0pt\lineskip.9pt
\ialign{$\mathsurround=0pt#1\hfil ##\hfil$\crcr#2\crcr\sim\crcr}}}

\newcommand{{\SD}}{\rm SD}

\newcommand{\vex}{\mbox{\boldmath${\rm x}$}}
\newcommand{\vey}{\mbox{\boldmath${\rm y}$}}
\newcommand{\ver}{\mbox{\boldmath${\rm r}$}}
\newcommand{\vesig}{\mbox{\boldmath${\rm \sigma}$}}

\newcommand{\veP}{\mbox{\boldmath${\rm P}$}}
\newcommand{\vep}{\mbox{\boldmath${\rm p}$}}

\newcommand{\vez}{\mbox{\boldmath${\rm z}$}}

\newcommand{\veL}{\mbox{\boldmath${\rm L}$}}

\newcommand{\veA}{\mbox{\boldmath${\rm A}$}}

\newcommand{\ven}{\mbox{\boldmath${\rm n}$}}

\newcommand{\veB}{\mbox{\boldmath${\rm B}$}}

\newcommand{\veE}{\mbox{\boldmath${\rm E}$}}
\newcommand{\veJ}{\mbox{\boldmath${\rm J}$}}
\newcommand{\veal}{\mbox{\boldmath${\rm \alpha}$}}
\newcommand{\vegam}{\mbox{\boldmath${\rm \gamma}$}}

\newcommand{\vePsi}{\mbox{\boldmath${\rm \Psi}$}}

\newcommand{{\Mc}}{\mathcal{M}}

\newcommand{\lan}{\langle}
\newcommand{\ran}{\rangle}

\begin{document}

\maketitle
\begin{abstract}\noindent
Relativistic Hamiltonians, derived from the path integrals, are
known to provide a simple and useful  formalism for hadrons
spectroscopy in QCD. The accuracy of this approach is tested using
the QED systems, and the  calculated spectrum is shown to
reproduce exactly that of the Dirac hydrogen atom, while the
Breit-Fermi nonrelativistic expansion is obtained using
Foldy-Wouthuizen transformation. Calculated positronium spectrum,
including spin-dependent terms, coincides with the standard QED
perturbation theory to the considered order $O(\alpha^s)$.
\end{abstract}

\section{Introduction}

 The path integral approach to QCD and QED was actively developed since the
 first formulation in \cite{1,2}  (see \cite{2*} for reviews, references,  and discussions). The particular line
 of development is the so-called Fock-Feynnman-Schwinger  Representation
 (FFSR) \cite{3,4,5}, where both relativism and gauge invariance are made explicit. In
 this latter framework one derives the  path-integral relativistic Hamiltonian to be called the Relativistic
  Hamiltonian (RH), originally exploited  in a simple form in \cite{6}.

  Recently a new integral form of the hadron Green's function and a rigorous
  derivation of the RH was done in \cite{6*}, and we
  shall use this latter form in what follows.

 As it is, the RH formalism is one of the most powerful methods in  QCD, which
 allows to predict spectra and wave functions of hadrons, using a minimal
 input: current quark masses, string tension, and $\Lambda_{QCD}$. Therefore it
 is very important to check its validity for different systems and the accuracy
 of results. In the course of  derivation some approximations have been done, the
 significance of which can be made clear by comparison with other relativistic
 approaches. In the case of  the one-particle system in  an  external field the
 basic approach is that of the Dirac equation and one can compare results of
 two approaches  -- the path integral  Hamiltonian and Dirac equation   in
 external fields, e.g. for the Coulomb case in QED.  In  the case of  the linear potential
 in QCD  results can be compared with lattice and experimental data.

 It is important  that the FFSR is derived for the Green's functions, and the
 RH appears in the kernel in the exponent and depends on
 additional integration variables, which play the role of virtual particle
 energies. Therefore one encounters the problem of the proper definition of the
 RH as an  operator  and its excited states. This topic will also
 be discussed in comparison with the Dirac formalism,  the  QED
 perturbation theory for positronium,  and relativistic quark
 models. As a result, we shall estimate the accuracy of
 approximations made and shall give the scheme of
 calculations for the  ground and excited states, both in one-particle and two-
 particle systems. As an additional topic, we compare nonrelativistic expansions for the RH with
 the known Breit-Fermi expansion.

 The plan of the paper is as follows. The
 short derivation of the RH is done in Section 2. Section 3 is devoted to the
 Breit-Fermi expansion of the RH. In Section 4  the spectrum of the RH for the
 hydrogen atom is compared with Dirac and Salpeter equations. The case of
 positronium and the accuracy of the spectrum of the RH is considered in
 Section 5. The last Section is  devoted  to the discussion of  results and
 perspectives.

\section{Derivation of the relativistic Hamiltonian}

 We start with the FFSR for the fermion propagator in the external gauge field
 $A_\mu$ in QED, as well as in QCD in the Euclidean space-time

 $$ S=(m+D)^{-1} = (m-\hat D) (m^2-\hat D^2)^{-1}=$$\be=(m-\hat D) \int^\infty_0 ds
 e^{-s(m^2-\hat D^2)}=(m-\hat D) \int^\infty_0 ds  (D^4z)
 e^{-K} W_F, \label{1}\ee

\be(D^4z)_{xy} = \int\frac{d^4p}{(2\pi)^4} \prod_k \frac{d^4\Delta
z(k)}{(4\Pi\varepsilon)^2}\exp [ip (\sum_k \Delta z (k) -
(x-y))],~~ N\varepsilon =s;\label{2}\ee where  the kinematic
kernel \be K= m^2s + \frac14 \int^s_0 \left(
\frac{dz_\mu}{d\tau}\right)^2 d\tau,\label{3}\ee  and the
generalized Wilson line  is \be W_F (x,y) = P_A \exp (ig \int^x_y
A_\mu dz_\mu+ g \int^s_0 \sigma_{\mu\nu} F_{\mu\nu} d_\tau),
\label{4}\ee where $P_A$ is the ordering operator in the case of
the nonabelian field $A_\mu$. Note, that matrices $\gamma_\mu$
enter $W_F$ only in the term \be \sigma_{\mu\nu} F_{\mu\nu} =
\left(
\begin{array} {ll}\vesig \veB&\vesig \veE_E\\\vesig \veE_E &
\vesig \veB\end{array} \right), \label{5}\ee
 and  here $\veE_E$  is  the Euclidean electric field, which should be replaced by $
 \veE_M \equiv i \veE$ in the Minkowskian case.  Hence all connections of large and small Dirac  components are provided by
 the electric field in $W_F$ and  the  factor $(m-\hat D)$ an (\ref{1}).

 As will be seen, the main difference between the RH and the  Dirac equation lies
 in two points:\begin{enumerate}
    \item  The RH is a quadratic operator, which stems from the quadratic
    combination $(m^2-\hat D^2)$, while the Dirac operator  is linear in
    momenta and fields $A_\mu$. This difference can be seen in the resulting
    nonrelativistic expansion of both operators and eigenvalues, and is  cured
    by the Foldy-Wouthuizen transformation with the  account of the factor
    $(m-\hat D)$ in (\ref{1}).

    \item  The  new  element in the relativistic path integral, as compared to
    its nonrelativistic analog,  is the time path in the quantum paths. As shown
    explicitly in \cite{6*}, the integration $ds(D^4 z)_{xy}$ in Eq. (\ref{1})
    can be  written,  using the relations  $$ s =\frac{T}{2\omega}, ~~ T  \equiv |x_4
    -y_4|,~~
    d \tau = \frac{dt_E}{2\omega}, $$ so that

 \end{enumerate}

 \be \int ds (D^4z)_{xy} e^{-K} W_F (x,y) = T \int^\infty_0
 \frac{d\omega}{2\omega^2} (D^3z)_{\vex\vey} e^{-K (\omega) } \lan \Phi_z
 (x,y)\ran_{\Delta z_4}, \label{6}\ee
 where \be K(\omega) =\int^T_0 dt_E  \left( \frac{\omega}{2} +
 \frac{m^2}{2\omega} + \frac{\omega}{2}  \left( \frac{ d\vez}{dt_E}
 \right)^2\right),
 \label{7}\ee
 and one can split the time element in (\ref{4}), $ dz_4 \to \Delta z_4= \Delta
 t_E + \Delta \tilde  z_4$, so that $\Delta t_E$ is  a monotonic Euclidean time
  interval, while $\Delta \tilde z_4$ is a stochastic one, with $\sum^N_{k=1}
  \Delta \tilde z_4 (k) =0$. Correspondingly, in the integral $(Dz_4)_{x_4y_4} =
  \int\frac{dp_4}{2\pi} \prod_k \frac{d\Delta z_4 (k)}{\sqrt{4\pi
  \varepsilon}}$ $\exp [ip_4 (\sum_k \Delta z_4(k) -T)]$ of the Wilson line
  (\ref{4}) one can write
  \be \lan \Phi_z (x,y) \ran_{\Delta z_4} =\int(Dz_4)_{x_4 y_4} W_F (x,y) =
  \overline{ W_F (x,y)} \sqrt{\frac{\omega}{2\pi T}} \varphi \{
  A_\mu\}.\label{8}\ee

  Here $\varphi \{ A_\mu =0\} =1$ and also $\varphi=1$ for $A_\mu$ independent
  of $z_4$.
As it was argued in \cite{6*}, the difference $(\varphi\{A\} -1)$ takes into
account the creation of additional particles and hence higher Fock components
in the  total wave function and higher Fock matrix elements in the Hamiltonian.
In terms of one - or  two- particle Green's   functions these contributions can
be considered as radiative corrections, which are absent in the simplest form
of Dirac or Bethe-Salpeter equation. In what follows we shall consider only the
minimal Fock component and use the condition $\varphi\{A\} \equiv 1$.

  Now the monotonic  part $\bar W_F$ depends only on  3d  trajectories
  $z_\mu =\{\vez (t_E), t_E\}$ and  is equal to \be \bar W_F (x,y) = \exp \left\{
  \int^T_0  dt_E \left[ ig A_4 (t_E) + ig A_i \frac{ dz_i}{dt_E} + g
  \frac{\sigma_{\mu\nu} F_{\mu\nu}}{2\omega}\right] \right\}.\label{10}\ee
  In what follows we shall test the approximation of smooth trajectories
  with $\varphi \{ A_\mu\} \equiv 1$  and compare  the corresponding results with
  exact calculations of Dirac equation for the Coulomb  potential.

  As a result, from (\ref{6}), (\ref{7}), (\ref{8}) one can write  the
  Hamiltonian  for a fermion in the electromagnetic field $\{ \veA (\vez, t),
  A_0 (\vez, t)\}$,\be H(\omega) = \frac{ (\vep - e
  \veA)^2}{2\omega} + \frac{m^2+ \omega^2}{2\omega} + eA_0 - \frac{e(\vesig
  \veB)}{2\omega} - \frac{ie ( \veal \veE)}{2\omega}, ~~ \veal = \left(
  \begin{array}{ll} 0&\vesig\\ \vesig&0\end{array}\right),\label{11}\ee

One can see, that the obtained Hamiltonian contains the parameter $\omega$,
which plays the role of the virtual particle energy, to be integrated over in
the expression (\ref{6}) for the Green's function. There are several ways to
proceed and  get the final spectra, which are discussed in what follows.
  In the next   Section  we compare   $H$   with nonrelativistic
    expansions  of the Dirac equation.

  \section{ Nonrelativistic expansions in RH and Dirac equation}

  We start with the Dirac equation for the hydrogen-like atom, where $A_0 =
  -\frac{Z\alpha}{r}$, and take into account  that the exact form of  the
  fermion Green's function (for $\varphi \{A\}\equiv 1)$  is
  \be G(x,y) =  \sqrt{\frac{T}{8\pi}} \int^\infty_0
  \frac{d\omega}{\omega^{3/2} }(m- \hat D)_x\lan \vex |e^{-H(\omega) T}\vey\ran.\label{13}\ee
  In (\ref{13}) the Hamiltonian was defined in Minkowskian space-time,  while the
  final expression is written for the Euclidean time  $T$  (see Appendix   in  \cite{6*} for
  details of  derivation).

  As the next step we consider  the ``projection operator'' $(m- \hat D)_x$ in
  the integral   (\ref{13}), and take into account that the derivative
  $\frac{\partial}{\partial x_\mu}$ is acting on the Wilson line (\ref{4}),
  resulting in the following expression (see Appendix 1 of the  second ref. in
  \cite{6}),
  \be (m-\hat D) \to \beta \left( \begin{array}{ll}  \omega+m& \vesig \vep\\
  \vesig \vep & \omega -m\end{array} \right).\label{14}\ee

  At this point one needs  to diagonalize the whole expression under the
  integral (\ref{13}), which allows to  give   the energy eigenvalues of
  the Hamiltonian $ H$  with the account of the  lower components of the wave
  function. In this way one writes
  \be (m-\hat D) = \beta U^+ \left( \begin{array}{ll} \omega + \sqrt{
  \vep^2+m^2}& 0 \\
0&  \omega - \sqrt{
  \vep^2+m^2}\end{array} \right) U, \label{15}\ee
  where \be U=  \left( \begin{array} {ll} \cos \theta & - \sin \theta\\
  \sin\theta& \cos \theta \end{array} \right) = e^{iS}; ~ tg 2 \theta =
  \frac{\vesig\vep}{m}, ~S= \frac{i\vegam \vep \theta}{p}= - \beta\alpha_2\theta. \label{16}\ee


In a similar way one can write \be H(\omega) = U^+ \tilde H (\omega) U, ~~
\tilde H(\omega) = e^{iS} H(\omega) e^{-iS}.\label{17}\ee

Our reasoning below and in the next Section follows the arguments from the book
\cite{7}, and $\tilde H(\omega)$ can be found as a series (see chapter 2 of
\cite{7}) \be \tilde H (\omega) = H(\omega) + i [S,H] -\frac12 [S,[S,H]]-...~~.
\label{18}\ee

As one can see in (\ref{16}), the series in (\ref{18}) is in powers of
$\left(\frac{p}{m}\right)$ and gives the higher orders of the nonrelativistic
expansion, whereas the first two orders are contained already in $H(\omega)$.
Indeed,  keeping for  simplicity  the first three terms in (\ref{11}), which we
denote as $H_0(\omega)$,

\be H_0 (\omega) = \frac{(\vep-e\veA)^2}{2\omega} +
\frac{{m^2+\omega^2}}{2\omega}+eA_0, \label{19}\ee and taking into account that
at large $T$ the integration over $d\omega$ in (\ref{13}) can be done   using
the minimum of $H_0(\omega)$ in momentum space  at some $\omega=\omega_0$, one
has \be \omega_0 = \sqrt{(\vep- e\veA)^2+m^2}, ~~ H_0 (\omega_0) = \omega_0
+eA_0.\label{20}\ee

Now the nonrelativistic expansion of $\omega_0$  and $ H_0(\omega_0)$ in powers
of $\left(\frac{p}{m}\right)$, yields  the first terms of the Breit-Fermi
expansion,
 namely, the so-called Pauli Hamiltonian \cite{8}, or to be more precise, its positive energy part.
Another root of $\omega_0, \omega_0 =-\sqrt{(\vep-e\veA)^2+m^2}$ is out of the
integration region, and  in the full Minkowskian integral   one would obtain
instead \be H_{+/-}(\omega_0) =\beta \sqrt{(\vep-e\veA)^2+m^2-e(\vesig\veB)}+ e
A_0. \label{21}\ee

We now turn to the next two terms in (\ref{18})  and take into
account that $\cos \theta \approx 1-\frac{p^2}{8m^2}, ~~ \sin
\theta \approx \frac{\vesig \vep}{2m}$, and hence, additional
terms from $eA_0$ and $\beta \vesig \veE$
  in (\ref{11}) yield \be U^+H(\omega)U=\cos \theta eA_0 \cos \theta+...=
\frac{e\Delta A_0}{8m^2} +...=-\frac{e div \veE}{8m^2}+...~~.\label{23}\ee In a
similar way in (\ref{23}) one obtains  the full $O(1/m^2)$ form,

$$ \tilde H  =\left( m+ \frac{(\vep-e\veA)^2}{2m} - \frac{p^4}{8m^3}\right)+ eA_0
- \frac{e}{2m} \vesig \veB+$$ \be +(-\frac{e}{4m^2} \vesig (\veE\times \vep))
-\frac{e}{8m^2 } div \veE, ~ \veE =-\nabla A_0 .\label{24}\ee

Note, however, that $div \veE\sim \delta^{(3)}(\ver)$ and the higher in
$\left(\frac{p}{m} \right)$ terms bring about even higher derivatives of  the $
\delta$-function, which  makes the evaluation of this Hamiltonian questionable.
Therefore it is more convenient
  from the beginning to consider the exact solution of the Dirac equation and
compare it with the exact eigenvalues of  $H(\omega)$ (\ref{11}), in this way
finding the accuracy of  approximations made in the path integral  method. This
is done in the next Section.

\section{Exact Dirac spectrum from RH for the hydrogen-like atoms}

Here we  study the energy eigenvalues of hydrogen-like atoms. From
(\ref{11}) the  RH is

\be H(\omega) = \frac{\vep^2}{2\omega} + \frac{m^2+\omega^2}{2\omega} + eA_0
-\frac{ie (\veal \veE)}{2\omega}.\label{25}\ee At this point one has two
possibilities:

 1) to calculate eigenvalues $M_n(\omega)$ of $H(\omega)$ and
then to find the stationary point $\omega_0$ of $M_n(\omega) $, yielding  the
actual eigenvalue $M_n(\omega_0)$.  This choice was used in \cite{6} and called
`` the einbein method'';

 2) to define $\omega=\omega_0$ from the condition
$\frac{\partial H(\omega)}{\partial\omega}|_{\omega=\omega_1}=0$,
finding in this way the ``stationary value'' of the  Hamiltonian.
This brings us to the (generalized) Salpeter equation \cite{11},
extensively studied in the framework of RH , e.g., in \cite{12}
and in the relativistic quark model \cite{13} (see \cite{14} for
reviews). In our case  $A_0 =- \frac{Z\alpha}{r}$  and
$\veE=-\nabla A_0 = \frac{Z\alpha}{r^2} \ven.$

We start with the simplest (einbein) procedure for the ground
state, solving the equation \be \left(
\frac{\vep^2}{2\omega}-\frac{Z\alpha}{r}\right)\psi = \varepsilon
\psi, ~ \varepsilon =- \frac{\omega (Z\alpha)^2}{2 n^2}, ~~
n=1,2...~~.\label{26}\ee Inserting $\varepsilon$ in (\ref{25}) and
neglecting there the last term on the r.h.s., one obtains the
expression  for the total
 eigenvalue $M_n(\omega)$:

\be H(\omega) \Psi_n = M_n (\omega) \Psi_n, ~~ M_n (\omega) =
\frac{m^2+\omega^2}{2\omega} - \frac{\omega (Z\alpha)^2}{2n^2}.\label{27}\ee

As it  is prescribed by the $\omega$ integration in (\ref{13}), the actual
 energy eigenvalue $M_n(\omega_0)$ should be obtained from $M_n (\omega)$ by the minimization
procedure: \be \left.\frac{\partial M_n (\omega)}{\partial \omega}
\right|_{\omega=\omega_0}=0; ~~ \omega_0 =
m\sqrt{1-\left(\frac{Z\alpha}{n}\right)^2}= M_n (\omega_0)\label{28}\ee This
form  should be compared with the exact Dirac Hamiltonian eigenvalues $M_n^D$
(see \cite{7}):

\be M^D_n =\frac{m}{\sqrt{1+\left(\frac{Z\alpha}{n-\delta_j}\right)^2}}, ~~
\delta_j=j+\frac12 - \sqrt{\left(j+\frac12\right)^2- (Z\alpha)^2}.\label{29}\ee

It is remarkable  that for the ground state with $n=1, j=\frac12$  the einbein
 approximation gives  exactly the same answer, i.e. \be M_1 (\omega_0) =
M^D_1\left(j=\frac12\right)\label{30}\ee

However, for higher levels the predictions of (\ref{28}) and
(\ref{29}) differ by $O(m(Z \alpha)^4)$. Moreover, $M_n(\omega_0)$
does not depend on $j$. In general, the einbein method gives a
reasonable approximation for not highly excited QCD bound states
\cite{12}, but in principle does not insure the orthogonality of
different wave functions. To overcome this, we turn to the second
possibility -- the square root or Salpeter equation.

 To  this end one keeps
 in $H(\omega)$    (\ref{11}) the last term for  the hydrogen-like
 atoms, $\veA=0, A_0 = - \frac{Z\alpha}{r}$, and
 (\ref{11})
 has  the form,

\be H(\omega) =\frac{\vep^2 +m^2 +\omega^2-ie\veal\veE}{2\omega} -
\frac{Z\alpha}{r},~~  H\Psi=M(\omega)\Psi.\label{36}\ee

As prescribed in the second ( square root or  ``Salpeter'') method, we define
$\omega$ from the minimum of the kinetic part, written in the momentum space,
  \be \left.\frac{\partial
H(\omega)}{\partial\omega}\right|_{\omega=\omega_0}=0;~~ \omega_0= \sqrt{
\vep^2 + m^2 - ie\veal \veE}.\label{38a}\ee Hence $H(\omega_0)$ acquires the
form \be \tilde H(\omega_0)=\sqrt{\vep^2+m^2-ie\veal\veE}-\frac{Z\alpha}{r},~~
\tilde H(\omega_0) \vePsi_n =\tilde M_n (\omega_0)\tilde \Psi_n.\label{39a}\ee
Notice that    in the chiral representation for $\gamma$ matrices one can write
$-ie\veal \veE \to \mp i Z\alpha \frac{\vesig \ven}{r^2}$.

To find the eigenvalues of $\tilde H(\omega_0)$ one can  write
$\sqrt{\vep^2 + m^2 -ie\veal \veE} \Psi_n = \left( \tilde M_n +
\frac{Z\alpha}{r}\right) \Psi_n $; multiply it with the Hermitian
conjugated equation  times $\beta$, \be \Psi^*_n \sqrt{\vep+ m^2+
i e \veal \veE} \beta \sqrt{ p^2+m^2-ie\veal \veE}\Psi_n =
\Psi^*_n \left(\tilde M_n + \frac{Z\alpha}{r} \right) \beta
\left(\tilde M_n + \frac{Z\alpha}{r} \right) \Psi_n,\label{32}\ee
obtaining in this way the Hamiltonian \be \left\{ \vep^2 + m^2 \mp
i Z \alpha \frac{\vesig \ven}{r^2} - \left( \tilde M_n +
\frac{Z\alpha}{r} \right)^2 \right\} \Psi_n =0. \label{33}\ee Then
following the same  procedure, as in \cite{7} for the same
Hamiltonian (see Appendix for details of derivation), one obtains
the exact Dirac spectrum (\ref{29}). In this way we arrived  at
the Dirac spectrum starting from the square root form (\ref{39a}),
using  the quadratic expression (\ref{33}).

However,    direct use of the square root form  in the $x$ space brings about
singularities around zero, as can be seen as follows. Indeed,  proceeding


$$ \sqrt{\vep^2 +m^2-ie \veal\veE}\Psi_n = \left(\tilde  M_n + \frac{Z\alpha}{r}\right) \Psi_n\to
$$\be \left(\vep+m^2-ie \veal \veE- (\tilde M_n + \frac{Z\alpha}{r})^2
\right) \Psi_n = X\Psi_n\label{40a}\ee with \be X=\left[ \sqrt{\vep^2 +m^2
-ie\veal \veE},~~ \frac{Z\alpha}{r}\right].\label{41a}\ee

One can see  that $X$ is a sum of  the  $\delta$-function and its derivatives.
These terms    can be neglected, if one excludes the small region
 around the  origin. It is  interesting  that  to solve Eq.(\ref{40a}) with $X=0$  one can use (\ref{33})
 with $\varepsilon_{\lambda, n}= -\frac{M_n(Z\alpha)^2}{2(n-\delta_j)^2}$,
and the resulting equation for $\tilde M_n$ is    \be \tilde M^2_n= m^2
-\frac{\tilde M_n^2 (Z\alpha)^2}{2(n-\delta_j)^2}, ~~ M_n =
\frac{m}{\sqrt{1+\left(\frac{Z\alpha}{n-\delta_j}\right)^2}},\label{42a}\ee
with $\delta_j$  given in (\ref{28}). Therefore    one obtains  again the exact
spectrum Dirac equation, if in the coordinate space one solves the square root
equation, excluding the near-zero region.

Notice, that  the case of the Coulomb potential in the square-root
(Salpeter-type) equation was studied  analytically in \cite{14*}, and a
singularity in the $S$- wave radial wave function $R_0 (r) \sim (m r)^{-\nu_0},
~\nu_0 \approx 0.086583$ was found there, while the spectrum was found in the
form ($l=0$) \be M_{n0} = \frac{2m}{\sqrt{1+ \alpha^2/4n^2}}.\label{36*}\ee
Note, that the term $(-ie \veal \veE)$ was not  present in \cite{14*}, and
hence $\delta_j$ does not enter in (\ref{36*}).

\section{Two-body QED  Hamiltonian from the path integral}

 For two-body systems there is no exact formalism to compare  with in QCD,
 since the Bethe-Salpeter equation is not operative with strong
 nonperturbative forces. In QED one can  use standard perturbation theory and
 Salpeter equation, which ensure very high accuracy of results. Our aim in this
 Section is to compare the RH spectrum for two oppositely charged particles
 (e.g. positronium) with the standard QED calculations.
We consider the problem of two charges $e_1$ and $e_2$ and write the two-body
Green's function without stochastic time contributions (radiative corrections)
as in \cite{6*,15} \be G_{e_1e_2} (x,y) = \frac{T}{2\pi} \int^\infty_0
\frac{d\omega_1}{\omega_1^{3/2}} \int^\infty_0 \frac{d\omega_2}{\omega_2^{3/2}}
\left(D^3z^{(1)} D^3z^{(2)}\right)_{\vex\vey} 4tr Y_\Gamma \lan W \ran \exp
(-K_1-K_2),\label{37}\ee where \be Y_\Gamma = \frac14 \Gamma_1 (m_1 - i\hat
p_1) \Gamma_2 (m_2 - i \hat p_2),\label{38}\ee \be K_i = \int^T_0 dt_E \left(
\frac{\omega_i}{2} + \frac{m^2_i}{2\omega_i} + \frac{\omega_i}{2} \left(
\frac{d\vez^{(i)}}{dt_E}\right)^2 \right).\label{39}\ee In (\ref{37}) the
function  $W$ is the vacuum averaged contour integral over paths of charges
$e_1$ and $e_2$  in the   e.m. field $A_\mu$\be W= \lan \exp \left(
\sum_{k=1,2} \left( e_k^i \int A_\mu (z^k)  dz_\mu^k + e_k \int^T_0
\frac{dt_E}{2\omega_k} (\sigma_{\mu\nu} F_{\mu\nu})\right)\right) \equiv
e^{-\hat V_T}.\label{40}\ee

In  the  case $e_1=-e_2$, $W$ is the gauge invariant QED analogue of the Wilson
loop, and  below we shall consider this case for simplicity. To get rid of the
c.m. motion one integrates over $d(\vex-\vey)$ and obtains $$ \int d^3
(\vex-\vey) G_{e_1e_2} (x,y) = \frac{T}{2\pi} \int^\infty_0
\frac{d\omega_1}{\omega_1^{3/2}} \int^\infty_0 \frac{d\omega_2}{\omega_2^{3/2}}
Y_\Gamma d^3(\vex-\vey) e^{i\veP (\vex-\vey)}\times$$

\be \times\lan \vex \left|~ e^{-H(\omega_1, \omega_2, \vep_1, \vep_2) T}\right|
{\vey}  \ran,\label{41}\ee

$$H(\omega_1, \omega_2, \vep_1, \vep_2) = \sum_i \frac{\vep^2_i + m^2_i +
\omega_i^2}{2\omega_i } + \hat V=$$\be= \sum_i \frac{  m^2_i +
\omega_i^2}{2\omega_i }+ \frac{\vep^2}{2\tilde \omega} + \hat
V+\frac{\veP^2}{2(\omega_1 + \omega_2)}, ~~ \tilde \omega =
\frac{\omega_1\omega_2}{\omega_1 + \omega_2}.\label{42}\ee
 Since the last  term on the r.h.s.  in (\ref{42}) vanishes, one is left with
 the c.m. Hamiltonian,
 \be H(\omega_1, \omega_2, \vep) = \sum_{i=1,2}
 \frac{m^2_i+\omega^2_i}{2\omega_i}  + \frac{\vep^2}{2\tilde \omega} + \hat V,
 \label{43}\ee
  where  the potential  $\hat V$ is to be found from the cluster  expansion of the Wilson
  loop. Keeping only the  $O(e^2)$ terms (bilocal correlators), one has (see
  \cite{15}, \cite{16} for details)\be \hat V = V_C(r) + \frac{(\vesig_1 \vesig_2V_4
  (r) + S_{12} V_3)}{12\omega_1\omega_2} +\left( \frac{\vesig_1 \veL}{ 4 \omega_1^2} +
  \frac{\vesig_2\veL}{4\omega_2^2} \right) \frac{1}{r} V_0'(r)+
  \frac{(\vesig_1+\vesig_2)\veL}{2\omega_1 \omega_2} \frac{1}{r} V_2'
  (r),\label{44}\ee
  where \be V_C(r) = \int^r_0 \lambda d \lambda \int^\infty_0  d\nu D^{(2)}
  (\lambda, \nu),\label{45}\ee
  \be V_4(r) =\int^\infty_{-\infty} d\nu \left(3D^{(2)} (r,\nu) +  2r^2
  \frac{\partial D^{(2)} (r,\nu)}{\partial r^2}\right),\label{46}\ee
   \be V_3(r) = - r^2 \frac{\partial}{\partial r^2} \int^{\infty}_{-\infty}
   d\nu D^{(2)} (r,\nu)\label{47},\ee
   \be V_0' (r) = r \int^\infty_0 d\nu D^{(2)} (r, \nu) , ~~ V_2'(r) = r
   \int^\infty_0 d\nu D^{(2)} (r,\nu), \label{48}\ee
   and $D^{(2)} (\lambda, \nu)$ is the quadratic correlator
   \be e^2 \lan  F_{\mu\nu} (x) F_{\lambda\rho} (y) \ran = \frac12 \left[
   \frac{\partial}{\partial u_\mu} (u_\lambda \delta_{\nu\rho} - u_\rho
   \delta_{\lambda\nu} )  + \left(\begin{array}{l} \mu\leftrightarrow
   \nu\\\lambda\leftrightarrow \rho\end{array}\right) \right] D^{(2)} (u)
   \label{49}\ee
   with  $u=x-y$. To the  lowest  order $D^{(2)}(u)$ is $(e_1=-e_2=e)$\be D^{(2)}
   (u) = \frac{4\alpha}{\pi u^4}, ~~ \alpha= \frac{e^2}{4\pi}.\label{50}\ee

   Note, that the accurate derivation of the spin-dependent terms,  valid both
   for QCD and QED, taking into account the proper positions of $(m_i -\hat
   D_i)$ terms, is  done in \cite{15}. In the QED case   substituting $D^{(2)}$ from (\ref{50})  one obtains   the familiar
   results $ ( S_{12} = \frac14 (3\vesig_1 \ven \vesig_2 \ven- \vesig_1
   \vesig_2))$
   \be V_C (r) =- \frac{\alpha}{r}, ~~ \frac{1}{r} V_0' = \frac{1}{r} V_2' =
   \frac{\alpha}{r^3}; ~~ V_3 = \frac{3\alpha}{r^3}, ~ V_4 = 8\pi\alpha
   \delta^{(3)}(\ver).\label{51}\ee

   These expressions coincide with the corresponding nonrelativistic
   spin-dependent potentials, when $\omega_i = m_i$, but in our case (\ref{44})
   (\ref{51})  are applicable in the relativistic case to the order
   $O(\alpha^5)$. Note, that in the case of positronium the additional term in
   $\hat V$ appears due to the annihilation diagram, which in the
   nonrelativistic limit is \be V_5 = \frac{\pi\alpha}{2m^2} (\vesig_1 \vesig_2 +3)\label{51a}\ee

   One can now proceed as in (\ref{36}), (\ref{38a}), but treating all terms in
   $\hat V$ (\ref{44}) as a perturbation,  except for $V_C(r)$,  and
   for  $m_1=m_2=m$, $\Delta \hat V\equiv \hat V - V_{C }(r)$ one obtains \be \tilde
   H_{e,-e} = 2 \sqrt{\vep^2+m^2} - \frac{\alpha}{r} + \Delta \hat V \equiv H_0
   + \Delta \hat V.\label{52}\ee
   Again, as in (\ref{33}), for $\Psi^{(0)}_n$, $H_0 \Psi_n^{(0)} =
   M_n^{(0)}\Psi_n^{(0)}$, one has \be \{ 4 (\vep^2 +m^2) - (M_n^{(0)} +
   \frac{\alpha}{r})^2\} \Psi_n^{(0)}=0\label{53}\ee
   and the analog of the angular operator $\hat N^2$ (see Appendix) is now
   diagonal with eigenvalues $\lambda(\lambda+1)= L(L+1) - \frac{\alpha^2}{4}$,
   yielding the eigenvalues $\varepsilon_n = - \frac{M_n^{(0)}
   \alpha^2}{8\tilde n^2}, ~~ \tilde n = n - \delta_L$, with \be \delta_L = L- \sqrt{
   (L+\frac12)^2 - \frac{\alpha^2}{4} } + \frac12.\label{54}\ee
   Finally one obtains for $M_n^{(0)}$,
   \be M_n^{(0)} = \frac{2m}{\sqrt{1+ \frac{\alpha^2}{4\tilde
   n^2}}}~.\label{55}\ee

   The expansion in $\alpha^2$ produces the expected result,
   \be M_n^{(0)} = 2m - \frac{\alpha^2m}{4\tilde n^2}+... \approx 2m -
   \frac{\alpha^2m}{4n^2}+O(\alpha^4).\label{56}\ee

At this point we can compare the accuracy of our expressions
(\ref{55}) with the account  of  the   potentials $V_4, V_5$ in
(\ref{50}), (\ref{51}) to the results of QED perturbation theory
for the  orthopositronium ($1^3S_1- 2^3S_1)$ interval $\Delta E$
(see reviews \cite{17,18} for results and discussions). From
\cite{17}, Table 5, one obtains in perturbation theory $\Delta
E_{PT} = \Delta E_{PT} (\alpha^2) +\Delta E_{PT} (\alpha^4)
+\Delta E_{PT} (\alpha^n,n\geq 5) $, where \be \Delta E_{PT}
(\alpha^2) = 1.2336907351\cdot 10^9 MHz,\label{57}\ee \be \Delta
E_{PT} (\alpha^4) = -82.0056\cdot 10^3 MHz, \label{58}\ee \be
\Delta E_{PT} (\alpha^5) = -1.5014\cdot 10^3 MHz.\label{59}\ee

At the same time our Eq. (\ref{55}) contributes the same amount in
the order $O(\alpha^2), ~~ \Delta E_{RH} (\alpha^2) = \Delta
E_{PT} (\alpha^2)$, while in $O(\alpha^4)$ its contribution from
$M_n^{(0)}$ is $\Delta E'_{RH} (\alpha^4) = 23.9515582 \cdot 10^3
MHz,$ and from the potentials $V_4, V_5$ one obtains $ \Delta
E^{\prime\prime}_{RH} (\alpha^4) = - 102.1933153 \cdot^3 MHz$, so
that the total contribution in the order $O(\alpha^4)$ is \be
\Delta E_{RH} (\alpha^4) \equiv \Delta E'_{RH} (\alpha^4) +\Delta
E_{RH}^{\prime\prime}(\alpha^4)  = -78.2417571 \cdot 10^3
MHz,\label{60}\ee
 which should be compared to $\Delta E_{PT} (\alpha^4)$, Eq. (\ref{58}). One
 can see, that the difference between these numbers is  of the order  $O(10^{-6})$ of the total result for $\Delta E$, and  is
 in the realm of the $O(\alpha^5)$  corrections. Note also, that the
 relativistic $O(\alpha^4)$ corrections, coming from the square root expression
 (\ref{55}), are of the  vital importance for the resulting accuracy. In this
 way we have proved, that  the square root  of  the two-body Hamiltonian (\ref{52}) is
 able to provide the high accuracy for  the positronium spectrum.

\section{Discussion of results}

We have  calculated  the spectrum of the  hydrogen-like atoms in QED, using our
RH, derived  in the framework of the path integral. This spectrum   exactly
coincides with the spectrum of the Dirac equation.

It was shown above, that in the first approach (the einbein
approximation), where the eigenvalues are  functions of virtual
energy $\omega$,  one obtains a reasonable   result for the
relativistic ground state energy, however for higher eigenvalues
corrections  are of the order of $ (Z\alpha)^4$.

At the same time the second approach, where the virtual energy is
defined on the operator level, provides the square-root- type
Hamiltonian, which yields the exact Dirac spectrum.  In this way
our results support the so-called Salpeter approach in the
relativistic quark models, which was so successful in predicting
hadronic states \cite{12,13,16,19}. However, in QCD  the string
correction needs to be   taken into account, to provide  orbital
and radial Regge trajectories \cite{12} in good agreement with
experiment.

We have also shown, how the Breit-Fermi nonrelativistic expansion
is obtained from our RH, when Foldy-Wouthuizen transformation is
applied.

Finally, the case of two oppositely charged particles was considered and all
interaction  terms, including spin-dependent ones, were derived and   included
in the resulting Hamiltonian. The latter contains both kinematic relativistic
effects and lowest order dynamic effects, and our formalism allows to
distinguish between two contributions. A short comparison to the standard QED
perturbation  results is done for the $(2^3S_1 - 1^3S_1)$ energy interval of
positronium  showing a good accuracy of the RH for the positronium spectrum.

Summarizing  these results,  one can consider RH as a reliable
tool for the studies  both in QED and of hadronic properties in
QCD with the proper comparison with lattice and experimental
results.

Another important line of development is the  theory of QED
systems in strong magnetic field, where the  RH approach was
formulated in \cite{6*, 15}, and  a new phenomenon of the magnetic
focusing is found in \cite{20}.

The author is greatly indebted to A.M.Badalian for useful
suggestions and criticism, and to M.A.Andreichikov, B.O.Kerbikov
and V.D.Orlovsky for discussions.

This paper was supported by the RFBR grant 1402-00395.

\newpage
\vspace{2cm}
 \setcounter{equation}{0}
\renewcommand{\theequation}{A \arabic{equation}}

\hfill {\it  Appendix  }

\centerline{\bf \large Explicit solution of  Eq. (33)}

 \vspace{1cm}

\setcounter{equation}{0} \def\theequation{A. \arabic{equation}}

Following \cite{7}  we  write Eq. (\ref{33}) in the form \be \{ -
\Delta_r + \frac{\hat N^2}{r^2} - \frac{ 2 Z \alpha}{r} M_n -
(M^2_n -m^2)\} \psi_n =0, \label{A1}\ee where $\hat N^2$ in the
chiral representation for the matrices $\alpha_i$ in the term $(-
i \veal \veE)$ is written in the diagonal form as $\mp \frac{i
Z\alpha (\vesig \ven)}{r^2}$. For the total angular momentum $\veJ
=\veL + \frac{\vesig}{2}$ with eigenvalues $j=\frac12,
\frac32,...$ one can define $\hat N^2$ as the matrix in the states
$l_{\pm}= j \pm \frac12$, which has the form \be \hat N^2 = \left(
\begin{array}{ll} l_+  (l_+ +1) - (Z\alpha)^2
& \mp i Z\alpha\\
\mp i Z\alpha & l_- (l_- + 1) - (Z\alpha)^2\end{array}\right).\label{A2}\ee The
eigenvalues of $\hat N^2 $ are found from (\ref{A2}) to be \be \hat N^2 =
\lambda (\lambda+1), ~~ \lambda = \sqrt{ \left(j+\frac12\right)^2 -(Z
\alpha)^2}-1, ~~\sqrt{ \left(j+\frac12\right)^2 -(Z \alpha)^2}, \label{A3}\ee
and writing $\lambda= \left(j\pm \frac12\right)- \delta_j$, one can define the
radial quantum number $n_r$, pertinent to $\Delta_r, n_r =0,1,2,...,$ and the
solution of the reduced Coulomb problem (the first three terms in (\ref{A1}))
is \be \varepsilon_n  = -\frac{(Z\alpha)^2 M_n}{2\tilde n^2},\label{A4}\ee
where $\tilde n = n_r+\lambda +1 = n_r + j\pm \frac12 +1 - \delta_j=
n-\delta_j, ~~ n=1,2,...$

Finally,  from (\ref{A1})  one finds that $M^2_n-m^2 = 2 M_n \varepsilon_n$, or
\be M_n = \frac{m}{\sqrt{1+\frac{(Z\alpha)^2}{(n-\delta_j)^2}}}.\label{A5}\ee

\end{document}